\documentclass[aps,orb,twocolumn,amssymb,showpacs]{revtex4-1}
\usepackage{graphicx}

\begin{document}
\def\be{\begin{equation}}
\def\ee#1{\label{#1}\end{equation}}

\title{Fermions in Brans-Dicke cosmology}
\author{L. L. Samojeden}\email{samojed@fisica.ufpr.br}
\affiliation{Departamento de F\'\i sica, Universidade Federal do
Paran\'a, Caixa Postal 19044, 81531-990 Curitiba, Brazil}
\author{F. P. Devecchi}\email{devecchi@fisica.ufpr.br}
\author{G. M. Kremer}\email{kremer@fisica.ufpr.br}
\affiliation{Departamento de F\'\i sica, Universidade Federal do
Paran\'a, Caixa Postal 19044, 81531-990 Curitiba, Brazil}

\begin{abstract}
Using the Brans-Dicke theory of gravitation we put under investigation a hypothetical universe filled with a fermionic field
(with a self interaction potential) and a matter constituent
ruled by a barotropic equation of state. It is shown that the fermionic field (in combination with the Brans-Dicke scalar field
$\varphi (t)$) could be responsible for a final accelerated era, after an initial matter dominated period.

\end{abstract}
 \maketitle

\section{Introduction}

 Einstein's  General Relativity (GR) is the usually invoked
theory for  describing the
evolution of the fundamental space-time variables in cosmological models\cite{Weinberg}. As an alternative to GR
we have the Brans-Dicke theory of gravitation\cite{Weinberg,Wald}, a scalar-tensor formulation that rules
the gravitational phenomena through the interplay between the metric tensor and a scalar field $\varphi$ that controls the intensity
of the gravitational constant $G$.
On the other hand, cosmological models that include   dark energy  are strong candidates for
explaining the present  regime  of positive acceleration of the universe\cite{Weinberg,Turk}. Going in that direction
one possibility  is to consider fermionic fields as gravitational
sources for those accelerated universes; these sources has been
investigated using several
 approaches, with results including exact solutions,
anisotropy-to-isotropy scenarios and cyclic cosmologies (see, for
example~\cite{Saha, Armendariz, Obukhov, DKR}).
Recently, these authors~\cite{DKR} proposed a cosmological model, based on GR, in a
dissipative Universe and showed that in a young universe scenario
the fermionic field produces a fast expansion where matter (included via a
barotropic equation of state) is created till it starts to
predominate and the initial accelerated period gives place to a
decelerated era. In
this case the fermionic field plays the role of the inflaton in an
early period and the role of dark energy for an old
universe; without the need of a cosmological constant term or a
scalar field\cite{DKR}. In an exclusive old universe scenario an initially matter
dominated period gradually turns into a dark (fermionic) energy
period when an accelerated regime rules  for the rest of
the system evolution. One important point  is that it was
considered a self-interaction term for the fermionic constituent, in
the form of a function that included the
Nambu-Jona-Lasinio potential as a particular case~\cite{NJL}).  Testing those properties in a
universe ruled by Brans-Dicke gravitation would be of interest and this
is the main focus of this work.
 The manuscript is
structured as follows. In section II we make a brief review  of the tetrad formalism used  to include fermionic fields in a dynamical
curved space-time.  The Brans-Dicke field equations
for an isotropic, homogeneous and spatially flat universe are
derived in section III where we  present  the analysis
of the different scenarios in which the fermionic constituent would, in principle,
answer  for the transition to accelerated periods.  Finally we display
our conclusions. The metric signature used is  $(+,-,-,-)$ and units
have been chosen so that $8\pi G=c=\hbar=1$.

\section{ Fermions and gravitation}

  When one tries  to include fermions in a gravitational model what must be taken into account is that the gauge
group of general relativity does not admit a spinor representation.
 The tetrad formalism  solves the problem\cite{Weinberg,Wald,Ryder,Bir}.  Following
the general covariance principle, the metric tensor $g_{\mu\nu}$ satisfies
\begin{equation}
 g_{\mu\gamma} =e^a_\mu e^b_\gamma\eta_{ab},\qquad a
=0,1,2,3
 \label {1}
\end{equation}
 where $e^a_\mu$ denotes the tetrad or  vierbein and $\eta_{ab}$ is the Minkowski metric tensor;
latin indices refer to a
 local  inertial frame  whereas Greek indices to the  general coordinates system.
 The next step  to couple the fermionic field to gravity is to construct an action for the model;
we start reminding that the Dirac lagrangian density in  Minkowski space-time is
\begin{equation}
L_D=\frac{\imath}{2}[
\overline\psi\gamma^a\partial_a\psi-(\partial_a\overline\psi)\gamma^a\psi]-m\overline\psi\psi-V,
\label{2}
\end{equation}
where $m$ is the fermionic mass,
$\overline\psi=\psi^\dag\gamma^0$
 denotes the adjoint spinor
field. $V$ is a function of $\psi$ and
$\overline\psi$,  and represents a fermionic self-interaction. The general covariance principle
imposes that the Dirac-Pauli matrices $\gamma^a$ must be replaced by
their generalized counterparts $\Gamma ^{\mu}=e^\mu_a\gamma^a$,
where the new matrices satisfy the  extended
Clifford algebra,  $ \{\Gamma^\mu,\Gamma^\nu\}=2g^{\mu\nu}.$
In a second step we need to substitute the ordinary derivatives by their covariant versions
\begin{equation}
\partial_\mu\psi\rightarrow D_\mu\psi= \partial_\mu\psi-\Omega_\mu\psi,\quad
\partial_\mu\overline\psi\rightarrow
D_\mu\overline\psi=\partial_\mu\overline\psi+\overline\psi\Omega_\mu,
 \label{3}
\end{equation}
where  the spin connection $\Omega_\mu$ is given by
\begin{equation}
\Omega_\mu=-\frac{1}{4}g_{\mu\nu}[\Gamma^\nu_{\sigma\lambda}
-e_b^\nu(\partial_\sigma e_\lambda^b)]\gamma^\sigma\gamma^\lambda,
\label{4}
\end{equation}
with $\Gamma^\nu_{\sigma\lambda}$ denoting the Christoffel symbol; the generally covariant Dirac lagrangian then
becomes
\begin{equation}
L_D=\frac{\imath}{2}[ \overline\psi\,\Gamma^\mu
D_\mu\psi-(D_\mu\overline\psi)\Gamma^\mu\psi]-m\overline\psi\psi-V.
\label{5}
\end{equation}

  The field equations are then obtained from the  total action
\begin{equation} S(g,\psi,\overline\psi)=\int \sqrt{-g}\,L_T\,d^4 x,\label{6}
\end{equation}
 where  $L_T=L_G+L_D+L_M$ is the total lagrangian density. $L_M$ is the lagrangian density of the matter field and
 $L_G $ is the free gravitational  lagrangian density. In the Brans-Dicke theory\cite{Weinberg} we have

\begin{eqnarray}
L_{G} = \{R e^{\alpha {\varphi}} - \alpha^2 \omega e^{\alpha {\varphi}}
(\nabla {\varphi})^2\}, \label{}
\end{eqnarray}
 where $R$ is the curvature scalar and ${\varphi}$ is an scalar  field that controls  the gravitational  constant ($G_N$)
intensity\cite{Weinberg};
$\alpha$  and $\omega$
are constants\cite{Weinberg, Turk}. $L_D$   is the Dirac lagrangian  density
 (\ref{5}).
 Using the variational principle we
  obtain then the Dirac equations for the spinor field  (and its adjoint) coupled to the gravitational field
\begin{equation}
\imath\Gamma^\mu D_\mu\psi-m\psi-{dV\over d{\overline\psi}}= 0,\quad
\imath D_\mu\overline\psi\,\Gamma^\mu+m\overline\psi+{dV\over
d\psi}= 0. \label{7}
\end{equation}
Analogously, the Brans-Dicke gravitational  dynamics emerges from the total action (\ref{6})
\begin{equation}
G_{\mu\nu}=-T_{\mu\nu},\label{8}
\end{equation}
 where
\begin{eqnarray}\nonumber
G_{\mu\nu}&=&e^{\alpha {\varphi}} \{R_{\mu\nu} - \frac{1}{2} g_{\mu\nu} R - \alpha^2( g_{\mu\nu} \Box {\varphi}
 - \nabla_{\nu} \nabla_{\mu} {\varphi})\\
&-& \alpha^2 \omega[\frac{1}{2}g_{\mu\nu}(\nabla {\varphi})^2 -
\nabla_{\nu}{\varphi} \nabla_{\mu}{\varphi}]\}, \label{}
\end{eqnarray}
and  $T_{\mu\nu} $ is the  energy-momentum tensor of the sources of the gravitational field:
$T^{\mu\nu}=T^{\mu\nu}_D+T^{\mu\nu}_M$. The symmetric form of the fermionic energy-momentum tensor  is given by
\begin{eqnarray}\nonumber
 T^{\mu\nu}_D&=&\frac{\imath}{4}\big[\overline\psi\Gamma^\mu
 D^\nu\psi+\overline\psi\Gamma^\nu D^\mu\psi
\\
 &-&D^\nu\overline\psi
 \Gamma^\mu\psi
 -D^\mu\overline\psi\Gamma^\nu\psi\big]
 -g^{\mu\nu}L_D.
 \label{9}
 \end{eqnarray}
These equations rule  the dynamics of a Brans-Dicke universe filled with a fermionic and matter sources.

\section{Brans-Dicke and Dirac Field dynamics}

 Friedmann-Robertson-Walker (FRW) metric mirrors the
homogeneity and isotropy properties of the universe. The space-time interval is usually written as
\begin{equation}
ds^2=dt^2-a(t)^2(dx^2+dy^2+dz^2), \label{11}
\end{equation}
where $a(t)$ is the cosmic scale factor.
In the FRW metric  the tetrad components (1) become
\begin{equation}
e_0^\mu=\delta_0^\mu,\qquad  e_i^\mu=\frac{1}{a(t)}\delta_i^\mu,
\label{15} \end{equation}
 \noindent and the Dirac matrices turn out to be
\begin{equation}
\Gamma^0=\gamma^0,\quad  \Gamma^i=\frac{1}{a(t)}\gamma^i, \quad
\Gamma^5=-\imath\sqrt{-g}\,\Gamma^0\Gamma^1\Gamma^2\Gamma^3=\gamma^5
\label{16},
\end{equation}
  from which the spin connection  (see eq. (4)) is obtained; the non-zero  components are
$ \Omega_i=\frac{1}{2}\dot
a(t)\gamma^i\gamma^0.$
 For this isotropic and homogeneous universe the fermionic field  becomes an exclusive function of time; therefore
the Dirac equations (\ref{7}) for a non-massive fermionic field become
 \be
 \dot\psi+{3\over2}H\psi+\imath\gamma^0{dV\over
 d{\overline\psi}}=0,\quad \dot{\overline\psi}+{3\over2}H\overline\psi-\imath{dV\over
 d{\psi}}\gamma^0=0.
 \ee{18a}

 Besides the Dirac field, we have considered as gravitational source a matter constituent that is ruled by a barotropic
equation of state, namely $\rho_m=b_m p_m$, where $b_m$ is the barotropic coefficient ($0\leq b_m\leq 1$); then the total
energy density $\rho$ is given as a sum of these individual contributions, i.e.,
$\rho=\rho_D+\rho_m=T_{00}$, where  $T_{\mu\nu }$ is the total energy-momentum tensor of sources. The energy density $\rho$ satisfies the
 conservation law $ \dot \rho+3\dot a/a(\rho+p)=0 $. Besides,
the non-vanishing components of the fermionic energy-momentum tensor follow from (12) yielding
 \be
 {{(T_D)}^0}_0=V= \rho_D,
 \ee{19a}
 \be
  {{(T_D)}^1}_1=V-{dV\over
 d{\psi}}{{\psi}\over2}-{{\overline\psi}\over2}{dV\over
 d{\overline\psi}}=- p_D,
 \ee{19}
 which are exclusive functions of $\psi$ and $\overline\psi$.
 Combining the Dirac and
Einstein equations  one can obtain
   an independent  conservation law for the energy
 density of the fermionic field\cite{DKR}. This implies into a  decoupled conservation equation for the
energy density of the matter  constituent, i.e.,
 \be
 \dot \rho_m+3H(\rho_m+p_m)=0,
 \ee{21}
 where $H=\dot a/a$ is the Hubble parameter.

 In order to analyze the cosmological solutions of our model
  we have to define  first some   sources properties. For the  fermionic field we consider
a self-interaction  potential $V$; this can be modeled as an exclusive function of the scalar
invariant $(\overline\psi\psi)^2$;  following  the Pauli-Fierz theorem\cite{NJL, DKR} we have
 \be
 V=\left[(\overline\psi\psi)^2\right]^n,
 \ee{23}
where  $n$  is a constant real number. A particular case of the  Nambu-Jona-Lasinio potential~\cite{NJL}
 is obtained when $n=1$. In fact, the fermionic field behavior can be classified according to the value of the exponent
 $n$. In particular, for $n<1/2$ the
 pressure of the fermion field is negative and it could represent (in a universe ruled by Einstein gravity)
 either the inflaton or a dark energy constituent\cite{DKR}.
 The fermionic  energy density   is  given by
 $ \rho_D=\left[(\overline\psi\psi)^2\right]^n$.
 On the other hand the second constituent, the matter field, is described, as mentioned in the precedent section, following a barotropic
 equation of state\cite{DKR}.
After some algebraic manipulation we can put the model dynamics in the following form:
\begin{eqnarray}
2\frac{\ddot{a}}{a}+H^2+\alpha^2\ddot{{\varphi}}+2\alpha H\dot{{\varphi}}+{\alpha^2\omega\over2}
\dot{{\varphi}}^2 \nonumber
\\ =e^{-\alpha{\varphi}}\Big[(2n-1)(\bar{\psi}\psi)^{2n}+p_m\Big],\label{r1}
\end{eqnarray}
\begin{eqnarray}\nonumber
\alpha^2\ddot\varphi&+&3\alpha H\dot\varphi={e^{-\alpha\varphi}\over 3+2\omega}\Big[V+\rho_m
\\&-&3(2n-1)(\bar{\psi}\psi)^{2n}-3p_m\Big],\label{r2}
\end{eqnarray}
\begin{eqnarray}
\dot{\rho_m}+3H(\rho_m + p_m)=0,\label{matter}
\end{eqnarray}
\begin{eqnarray}
\dot\psi+{3\over2}H\psi+\imath\gamma^0{dV\over
 d{\overline\psi}}=0,\quad \dot{\overline\psi}+{3\over2}H\overline\psi-\imath{dV\over
 d{\psi}}\gamma^0=0.\label{psi}
\end{eqnarray}

We consider a pressureless matter field $(p_m=0)$, so that one can obtain from equation (\ref{matter}) that $\rho_m(t)=\rho_m(0)/a(t)^3$. Furthermore, the equations (\ref{psi}) can also be integrated for the potential (\ref{23}) resulting that the fermionic bilinear evolution is governed  by $[\overline\psi\psi](t)=[\overline\psi\psi](0)/a(t)^3$.
The remaining equations (\ref{r1}) and (\ref{r2}) constitute a highly non-linear system of differential equations and we proceed to solve it numerically.
 We  analyze  the time evolution of
our model choosing first
the conditions  for $t=0$:
 \begin{eqnarray}\nonumber
  a(0)&=&1, \quad\dot a(0) = 1, \quad [\bar \psi \psi](0)=0.001,\quad \rho_m(0)=1,\\\nonumber
 \varphi(0)&=&1,\quad \dot\varphi(0)=0.001.
 \end{eqnarray}
These conditions  characterize qualitatively an
initial proportion between the constituents; an era  when matter predominates over the fermionic density.
Besides that, we
 have to specify   the magnitude of the remaining parameters:
 we suppose initially that $ \alpha=1.0$, $\omega= 4\times10^4$ \cite{Will}  and  a value for the potential power $n$,  $n=0.2$ \cite{DKR}. 
 
 These choices are reference values that permit final adjustments to follow  several cosmological
constraints, like the   spectrum  of the Brans-Dicke coupling $\omega$\cite{Will}   and the present value of the Brans-Dicke scalar 
field $\varphi$.  These parameters can be in fact adjusted  due to  invariance properties of the Brans-Dicke gravitational field 
equations, by  using the following change of variables:
$\alpha \rightarrow \bar \alpha,\quad  \varphi \rightarrow \bar \varphi, \,\,\,\,\ \bar \alpha \bar \varphi = \gamma / \varphi + \alpha $
where $ \exp(-\gamma) = x_0$ is associated to the asymptotic  value of $\bar \varphi _0 = (2\omega +4)/(2\omega +3)$. These 
transformations show how a  value of the Brans-Dicke  parameter $\omega$ is linked to the definition of a new Brans-Dicke scalar field $\bar \varphi$.  
In fact, after numerical integration,  it is possible to verify that these features  are included in the $\varphi$ evolution (see figure 1) that imply  into the evolution of the gravitational ``constant" as  $G(t)=(2\omega + 4)/[(2\omega +3) \bar \varphi (t)]$.

 In figure 2 it is plotted the acceleration field $\ddot a$ as function of
 time $t$ for two different values of n. The results show that initially the universe is expanding
with negative acceleration, a period where the matter constituent predominates over the fermionic field.
With the evolution of time we have  increasing values in energy transference  to the fermionic field (this is not happening   directly but via the gravitational
field $a(t)$ and the scalar field $\varphi(t)$, as the equations of motion (20-23) show). The negative pressure of $\psi(t)$ help in fact to promote a final accelerated period (also shown  in figure 1)  indicating that in this model the dark energy  role would be played by a {\it combination of the fermionic constituent with the scalar}  $\varphi(t)$.
Other interesting results appear  when we choose the power $n$ to be in the neighborhood  of values
 $n\approx 0.33$. In fact, for a fixed value of $\dot \varphi (0)$ and increasing  values of $n$ what emerges is a universe that fails to show a final accelerated period, even when the fermionic field still exhibits a negative pressure ($
 p_D=(2n-1)\left[(\overline\psi\psi)^2\right]^n$). On the other hand, for values $n\leq 0.33$ we will find a universe that is permanently in accelerated expansion. Another important remark here is that this general qualitative behavior  depends strongly on the initial value of
the time derivative of  the scalar field $\varphi(t)$:   what we verify is that increasing values of $\dot \varphi (0)$ promote an earlier entrance on the accelerated period.

In figure 3 it is
 shown the behavior of the energy densities of the fermionic field $\rho_D$ and matter $\rho_m$
 as functions of time $t$. The numerical results show that eventually the
 energy density of the femionic field overcomes the energy density of the
 matter field, although this does not coincide with the instant    when  the universe goes into an accelerated period (opposed to what  occurs in some Einstein gravity based models\cite{KD}).   Another feature showed by the numerical results is that for larger values of  $n$ it follows that: (a) the energy density of fermionic field grows more slowly causing
 a larger decelerated period and (b) the energy density of the matter
 field has a less significant decay and also promotes a larger decelerated period.
 On the other hand, as  figure 3 shows, both densities have decreasing  values in time
  due to the permanent  expansion of the  universe (this was verified with plots of the scale factor against time).
Again, all cases are strongly
dependent on the exponent $n$  of the self-interacting potential and
one can obtain different behaviors,  that are in tune with the acceleration patterns presented above. Finally we verify that the
scalar field $\varphi$ approaches a final  constant value that  validates the accord between Brans-Dicke and Einstein gravitation  for large $t$.

As concluding remarks we stress that we have investigated the role of a fermionic field -- with a self-interacting
 potential  -- in an old Brans-Dicke universe. We have shown that  the fermionic
 field, in combination with the scalar $\varphi$  behave like  a dark
 energy constituent, promoting a final accelerated period after an initial era dominated by matter. These results depend strongly on
the initial values of $\dot \varphi$ and on the power $n$ present in the self-interactive fermionic potential.


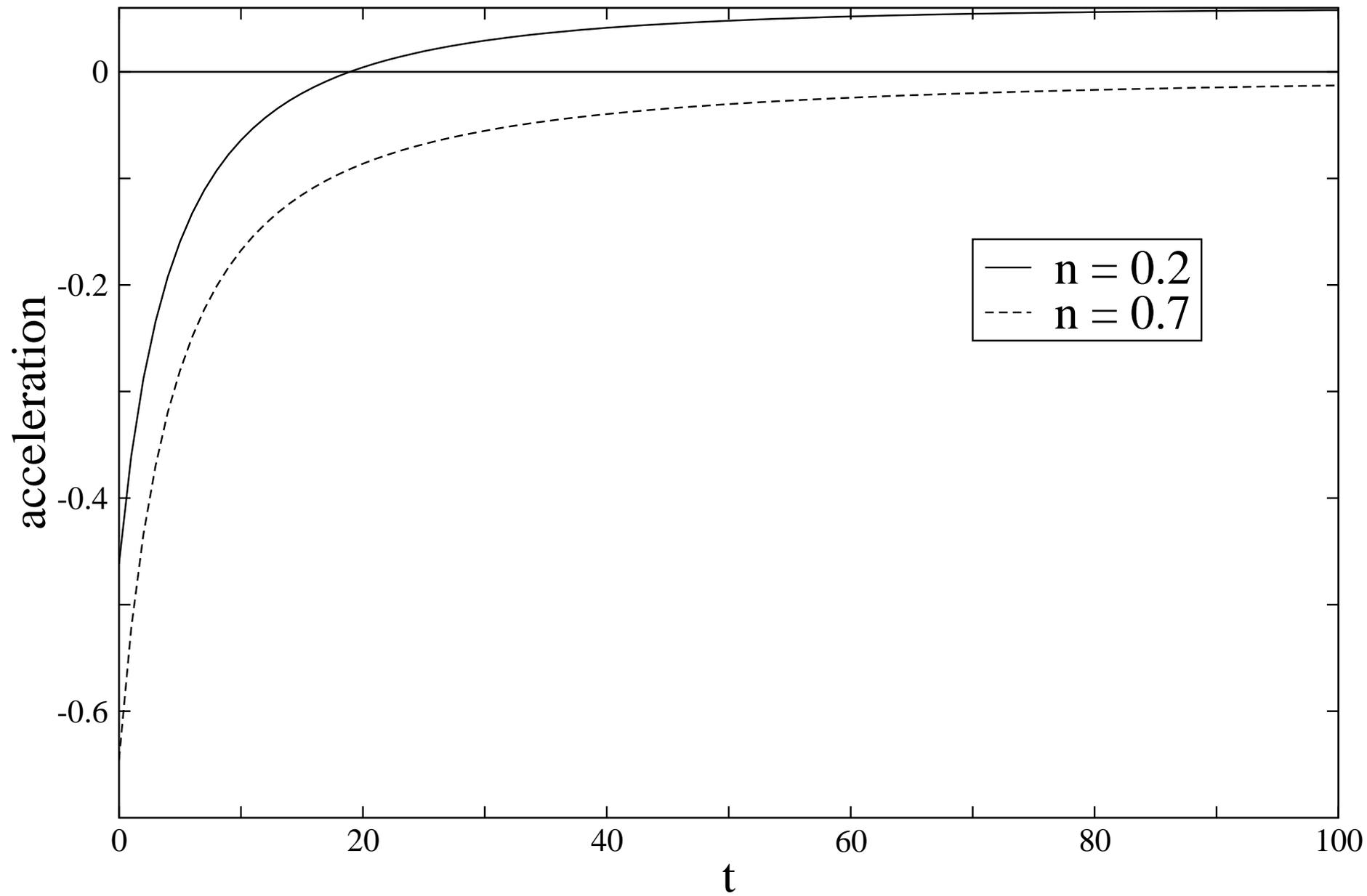
\end{document}